\documentstyle[12pt]{article}

\renewcommand{\baselinestretch}{1.62}

\begin{document}

\title{\bf Optimized perturbation theory for bound states:
toy model and realistic problem.}

\author{
{\bf A.A.Penin} {\it and} {\bf A.A.Pivovarov}\\
{\it Institute for Nuclear Research of the Russian Academy of
Sciences,} \\
{\it Moscow 117312, Russia}
}

\maketitle

\begin{abstract}
Within quantum mechanics model we study the problem of
resummation of an asymptotic perturbation series for
bound state parameters via optimization of the
perturbative expansion. A possible application of the method
to the positronium lifetime calculation is also
briefly considered. \\[0.5cm]
PACS number(s): 11.10.Jj,  11.10.St
\end{abstract}

\thispagestyle{empty}
\newpage


\section{Introduction}

A discrepancy between the experimental result \cite{1} and
theoretical predictions \cite{2} for the
orthopositronium width found quite a time ago  still persists in
spite of rigorous efforts to improve both theoretical
computations and experimental data. By now an agreement on
theoretical estimates for the width in the
next-to-leading order is obtained and this agreed estimate
is smaller than the experimental number by three standard
deviations that
causes some discomfort because it forces to allow quite big
contribution of higher orders.  The difference with
experiment could be accounted for if future calculations of
the order $(\alpha /\pi)^2$
coefficient determine it to be of order $250\pm 40$ \cite{1}.  The
anomalously large next-to-leading correction can appear
accidently in a sense that in higher orders corrections
become small but it can also be a signal of  bad convergence
of the series in fine structure constant for the positronium bound
energy (width). This does not seem  impossible since in most of
physically interesting models of quantum field theory
the conventional perturbation theory forms the asymptotic
series in coupling constant that can be used for
calculation of the Green's functions only if the effective
parameter of the expansion is small enough.  However when
the exact solution is absent it is difficult to determine
if the asymptotic expansion is applicable to compute some
physical quantity  and the value $\alpha =1/137$ can
be too large for the expansion in $\alpha $
being a good instrument in study of the orthopositronium
width though it allows to compute another quantities, for
example, the electron anomalous magnetic moment with high
precision. Therefore it is instructive to try to go beyond
the asymptotic expansion and improve the
ordinary perturbation theory for orthopositronium width.

Some methods have been suggested to improve a convergence
of the conventional perturbation theory. The basic idea of
the optimized $\delta$ expansion  is to introduce
the artificial parameter $\delta$ which interpolates between
the theory we intend to solve with Hamiltonian $H$, and
another theory, with Hamiltonian $H_0(\lambda)$ ($\lambda$
is a set of auxiliary parameters not present in the original
theory), which is soluble and reflects the main properties
of the theory we are interested in.  One defines a
new Hamiltonian depending on $\delta$

\begin{equation}
H_\delta =H_0(\lambda )+\delta (H-H_0(\lambda )).
\label{1}
\end{equation}
Then any desired quantity is evaluated as  a perturbation
series in $\delta$, which is set equal to unity at the end of
the calculations. The convergence of the series is achieved
by optimization procedure \cite{3} {\it i.e.} by fixing
the parameters $\lambda$ at every finite order of the
expansion according to principle of minimal sensitivity
(PMS) at the point where the result is least sensitive to
their variation or principle of fastest apparent
convergence (FAC) at the point where the next term in the
series vanishes or somewhat else. Though above procedure is not
rigorous it gives good numerical results in most of the cases.  The
method has been mostly advanced in studying the anharmonic oscillator
\cite{4} where the convergence has been rigorously established
\cite{5}.

Thus it seems instructive to optimize the perturbation
theory in analysis of the positronium bound state.
However the calculation of the correction  to the
orthopositronium width is very involved even in the case
of ordinary $\alpha $ expansion. Therefore it is useful
to consider the simplest model that, nevertheless, retains most
relevant features of the real problem and, as we hope, can help to
gain some intuition to cure the difficulties with positronium.

\section{The model}

The problem we will study is the ground state in spectrum of the
stationary Schr\"{o}dinger equation in three dimensions

\begin{equation}
\left(-\Delta + U({\bf r})\right)\psi({\bf r}) =
E\psi({\bf r})
\label{2}
\end{equation}
where we imply $2m=\hbar=1$
and

\begin{equation}
U({\bf r})=
\left\{
\begin{array}{cc}
-\left({\pi^2\over 4}+\alpha\right)+\alpha^2r,~~~r<1,\\
 0,~~~~r>1.\\
\end{array}\right.
\label{3}
\end{equation}
At $r<1$ the potential~(\ref{3}) consists of two parts:
one is a constant and another depends linearly  on $r$.
The Schr\"{o}dinger equation with the
constant part of the potential only has bound states for any positive
$\alpha$ with the ground state energy determined by the
equation

\begin{equation}
\sqrt{-E_0}=\sqrt{{\pi^2\over 4}+\alpha}
\ \cot\!\left(\sqrt{{\pi^2\over 4}+\alpha+E_0}\right).
\label{6}
\end{equation}
Therefore if $\alpha$ is
small enough we can search for the ground state
energy of eq.~(\ref{2}) using perturbation theory and consider the
constant part of the potential to be responsible for creation of the
bound state while the linear term is a perturbation because
it is suppressed by an extra power of $\alpha$ ( the
parameter $\alpha$ should not be confused with the fine
structure constant).

On the other hand the exact solution of eq.~(\ref{2}) is
known and leads to the equation for the ground state energy

$$
\sqrt{-E}=\alpha^{2/3}
\left({{\rm Bi}(\xi_0){\rm Ai}'(\xi_1)-{\rm
Ai}(\xi_0){\rm Bi}'(\xi_1)\over
{\rm Bi}(\xi_0){\rm Ai}(\xi_1)-{\rm
Ai}(\xi_0){\rm Bi}(\xi_1)}\right),
$$
\begin{equation}
\xi_0=-\alpha^{-4/3}\left({\pi^2\over 4}+\alpha+E\right),~~~~
\xi_1=-\alpha^{-4/3}\left({\pi^2\over 4}+\alpha-\alpha^2+E\right)
\label{4}
\end{equation}
where Ai and Bi are Airy functions \cite{mh}.
Using the asymptotic expansion
of Airy functions at large negative $\xi_i$ (small $\alpha$) we
obtain an asymptotic series for the ground state energy

$$
E(\alpha)\sim \tilde E(\alpha)\equiv -{\alpha^2\over 4}
\left(1-\left({3\over
2}+{2\over\pi^2}\right)\alpha+\left({21\over\ 16}-{11\over
6\pi^2}+{13\over\pi^4}\right)\alpha^2-\right.
$$
\begin{equation}
\left. -\left({39\over 32}-{41\over 8\pi^2}+{35\over
6\pi^4}+{26\over\pi^6}\right)\alpha^3+\ldots\right).
\label{5}
\end{equation}
Substituting numerical values for the coefficients of the
expansion~(\ref{5})

\begin{equation}
\tilde E(\alpha)= -{\alpha^2\over 4}
\left(1-1.7026\alpha+ 1.2602\alpha^2-
0.7864\alpha^3+\ldots\right)
\label{num}
\end{equation}
we find that the series merely reveals bad convergence
near the point $\alpha\sim 1$. In fact the series diverges for any
positive $\alpha$ because the coefficients of the
expansion~(\ref{5}) in high orders grow factorially. That reflects
the presence of a singularity of the function $E(\alpha)$ at the
origin of the complex $\alpha$ plane. The form of the singularity can
be found directly from eq.~(\ref{4}):  $E(\alpha)$ has a cut along
negative semiaxis and a branching point at $\alpha =0$.  For
sufficiently small $|\alpha|$ it is an analytical function for
$-\pi<{\rm arg}\alpha<\pi$ therefore  the series~(\ref{5}) is Borel
recoverable \cite{as} {\it i.e.} we can extract complete information
on the function $E(\alpha)$ from its asymptotic expansion. Presence
of the singularity in the Green's function of eq.~(\ref{2})  reflects
the fact that at $\alpha =0$ the spectrum of eq.~(\ref{2}) changes
qualitatively and a discrete part of the spectrum appears.

Such a singularity  does not ultimately
lead to divergence of the expansion of a bound state energy.  For
example, the ground state energy of the Schr\"{o}dinger equation with
the constant part of the potential (\ref{3}) is expanded in the
convergent series for any finite positive $\alpha$
$$
E_0(\alpha)=-{\alpha^2\over 4}
\left(1-\left({1\over 2}-{2\over\pi^2}\right)\alpha+\left({5\over\
16}-{11\over 6\pi^2}-{3\over\pi^4}\right)\alpha^2-\right.
$$
\begin{equation}
\left. -\left({7\over 32}-{13\over 8\pi^2}-{3\over
2\pi^4}-{6\over\pi^6}\right)\alpha^3+\ldots\right).
\label{7}
\end{equation}
So there is no implicit singularity in the series for $E_0(\alpha)$.
However after inclusion of the perturbation such a singularity
appears and the full series~(\ref{5}) becomes divergent. This, in a
sense, simulates the positronium bound state where the
non-relativistic Coulomb potential that is taken to build the leading
order bound state Green's function does not lead to implicit
singularity of bound state energy in the fine
structure constant while the
relativistic corrections result, as we suppose, in the divergent
series.  Thus we can use our quantum mechanics analog as a test model
for further analysis of the positronium.

\section{Optimized expansion}
Our purpose now is to develop the optimized perturbation theory (OPT)
for our toy model.  Following  the general idea we have to choose the
form of the "unperturbed" Hamiltonian. The Hamiltonian

\begin{equation}
H_0(\alpha ')= -\Delta + U_0(\alpha ',~{\bf r})
\label{8}
\end{equation}
where

\begin{equation}
U_0(\alpha ',~{\bf r})=
\left\{
\begin{array}{cc}
-\left({\pi^2\over 4}+\alpha '\right),~~~r<1,\\
 0,~~~~r>1,\\
\end{array}\right.
\label{9}
\end{equation}
is the potential of the spherical well with changeable depth
seems to be the most appropriate choice. Here the set of
parameters $\lambda$ in eq.~(\ref{1}) is reduced to the single
parameter $\alpha '$ characterizing the depth of the
well. The ordinary perturbation theory corresponds to
$\alpha '=\alpha$ and $\delta =1$. Making the expansion in $\delta$
and setting $\delta =1$ we obtain in $n$-th order the series for the
ground state energy

$$
E_n(\alpha ,\alpha ')=E^{(0)}(\alpha ')+
E^{(1)}(\alpha ,\alpha ')+\ldots +E^{(n)}(\alpha ,\alpha '),
$$
$$
E^{(0)}(\alpha ')=-{\alpha '^2\over 4}
\left(1-\left({1\over 2}-{2\over\pi^2}\right)\alpha '+\left({5\over\
16}-{11\over 6\pi^2}-{3\over\pi^4}\right)\alpha
'^2+\ldots\right),
$$
\begin{equation}
E^{(1)}(\alpha ,\alpha ') =
{(\alpha '-\alpha )\alpha ' \over 2}\left(1-\left({3\over
4}-{3\over\pi^2}\right)\alpha '+\ldots\right)+
\label{10}
\end{equation}
$$
+ {\alpha^2\alpha '\over 4}\left(1+{4\over\pi^2}-\left({3\over
4}-{2\over\pi^2}+{12\over\pi^4}\right)\alpha '+\ldots\right),
$$
$$
E^{(2)}(\alpha ,\alpha ')=\left({(\alpha '-\alpha )\over
2}+\alpha^2\left({1\over 4}+{1\over\pi^2}\right)\right)^2+\ldots .
$$
where the explicit expressions for $E^{(i)}$ are expanded in
$\alpha '$ and $\alpha$.  In general if $\alpha '-\alpha =O(\alpha^2)$
the effective parameter of the expansion is proportional to

\begin{equation}
{\langle 0|U(\alpha )-U_0(\alpha ')|0\rangle
\over E_0(\alpha ')}\sim\alpha .
\label{145}
\end{equation}
So $E_n(\alpha ,\alpha ')$ after expansion in
$\alpha$ correctly reproduces the first $n$ terms of
eq.~(\ref{5}).  Thus if we are interested in only the asymptotic
expansion the choice of the start approximation does not play a
crucial role. If, however, we intend to go beyond the asymptotic
expansion we have to choose the "unperturbed" Hamiltonian to provide
the best convergence of the expansion. The most transparent and
conventional way
to optimize the expansion~(\ref{10}) is to
fix the  parameter $\alpha '$ in $n$-th order according to
PMS (FAC) at the value $\alpha^{PMS}_n$ ($\alpha^{FAC}_n$)
so that

\begin{equation}
{\partial E_n\over\partial\alpha '}{\bigg |}_{\alpha '= \alpha^{PMS}_n}
=0,
\label{11}
\end{equation}
or

\begin{equation}
E^{(n+1)}|_{\alpha '= \alpha^{FAC}_n}
=0.
\label{12}
\end{equation}
At $n=1$ we have

\begin{equation}
\alpha^{PMS}_1=\alpha \left(1-\left({1\over 2}+{2\over\pi^2}\right)
\alpha+\ldots\right) ,
\label{13}
\end{equation}
\begin{equation}
\alpha^{FAC}_1=\alpha \left(1-\left({1\over 2}+{2\over\pi^2}\right)
\alpha+\ldots\right).
\label{14}
\end{equation}
Since $\alpha^{\#}-\alpha =O(\alpha^2)$ where $\#$ stands
for PMS or FAC we have $E^{(n)}(\alpha ,\alpha^{\#})=$ \\
$O(\alpha^nE^{(0)}(\alpha))$. Therefore  expanding
eq.~(\ref{11},~\ref{12}) in $\alpha$ we obtain at
$n\rightarrow\infty$ the formal series

\begin{equation}
\alpha^{\#}_n\rightarrow\alpha^{\#}=
\alpha (1+a^{\#}_1\alpha+a^{\#}_2\alpha^2+\ldots),
\label{15}
\end{equation}
moreover

\begin{equation}
E_n(\alpha ,\alpha_n^{FAC})|_{n\rightarrow\infty}
\rightarrow E_0(\alpha^{FAC}(\alpha)).
\label{155}
\end{equation}

Now we  show that one can choose
$\alpha^{PMS}=\alpha^{FAC}$. Let us suppose that
$a^{PMS}_i=a^{FAC}_i$ for $i\le n$. The PMS condition to
determine $a^{PMS}_{n+1}$ reads

\begin{equation}
{\partial E_{n+1}\over\partial\alpha '}{\bigg |}_
{\alpha '-\alpha^{PMS}=O(\alpha^{n+3})}=
O(\alpha^{n+1}E_0(\alpha)).
\label{16}
\end{equation}
On the other hand

$$
E^{(n+1)}|_{\alpha '-\alpha^{PMS}=O(\alpha^{n+2})}
=O(\alpha^{n+2}E_0(\alpha)),
$$
\begin{equation}
\tilde E(\alpha)-E_{n+1}|
_{\alpha '-\alpha^{PMS}=O(\alpha^{n+2})}
= O(\alpha^{n+2}E_0(\alpha))
 \label{17}
\end{equation}
because  $a^{PMS}_i=a^{FAC}_i$ for $i\le n$.
Therefore

$$
{\partial E_{n+1}\over\partial\alpha '}{\bigg |}_
{\alpha '-\alpha^{PMS}=O(\alpha^{n+3})}=
{\partial (E_{n+1}-\tilde E(\alpha))\over\partial\alpha '}{\bigg |}_
{\alpha '-\alpha^{PMS}=O(\alpha^{n+3})}=
$$
\begin{equation}
={\partial (E_{n+1}|_{\alpha '-\alpha^{PMS}=O(\alpha^{n+2})}
-\tilde E(\alpha))\over\partial\alpha '}
+O(\alpha^{n+1}E_0(\alpha))= O(\alpha^{n+1}E_0(\alpha))
\label{18}
\end{equation}
where taking a derivative of the asymptotic series is  justified
because $\tilde E(\alpha)$ is a power series and all non-analytical in
$\alpha '$ terms in $E_{n+1}$ must be cancelled in higher orders
and, therefore, can be omitted before taking a derivative.  So
condition~(\ref{16}) is satisfied for arbitrary $a^{PMS}_{n+1}$ and
we can put $a^{PMS}_{n+1}=a^{FAC}_{n+1}$. Since $a^{PMS}_i=a^{FAC}_i$
for $i=1$ (see eqs.~(\ref{13},~\ref{14})) one can choose
$a^{PMS}_i=a^{FAC}_i$ for any $i$ whence
$\alpha^{PMS}=\alpha^{FAC}$.

Thus in the limit  $n\rightarrow\infty$ all
corrections in the optimized expansion~(\ref{10}) vanish
both for PMS and FAC optimization prescription
and we obtain

\begin{equation}
E_n(\alpha ,\alpha_n^{\#})\rightarrow E_0(\alpha^{\#}
(\alpha))=E(\alpha).
\label{19}
\end{equation}

Since the function $E_0(\alpha^{\#})$ can be expanded in  the
convergent series in $\alpha^{\#}$ (eq.~(\ref{7})) the
singularity of the function $E(\alpha )$ at $\alpha =0$ is
absorbed by the function $\alpha^{\#}(\alpha)$ {\it i.e.}
the series~(\ref{14}) is a divergent asymptotic expansion.
However we can search for the values $\alpha_n^{\#}$ which
satisfy eqs.~(\ref{11},~\ref{12}) at every order
numerically rather than as a series in $\alpha$.
In Fig.~1 and Fig.~2 we plot the functions  $E_1(1.0,\alpha ')$
and $E^{(1)}(1,\alpha ')/E^{(0)}(1,\alpha ')$ to show a typical
pictures that are used to determine $\alpha^{\#}$. Thus for
$\alpha =1.0$ we obtain $\alpha_1^{PMS}=0.310$ and $\alpha_0^{FAC}=
0.304$.

Thus we have three  kinds of perturbative expansion: the asymptotic
series, the standard perturbation theory and OPT. The asymptotic
series seems to be the most primitive tool and we expect to obtain
the best results using OPT. This assumption is completely
confirmed by numerical analysis given in the next section.

\section{Numerical evaluation}

The results of numerical analysis are given in
Tabs.~$1-3$\footnote{For the exact value and  approximations of
the bound energy in Tabs.~$1-3$ the factor $(-10^{-2})$ is implied}.
In Tab.~1 we establish  the exact value $E(\alpha)$ along with the
results of the asymptotic expansion $\tilde E_n(\alpha)$ up to the
$n$-th  order ($n=0,~1,~2,~3$). In Tab.~2 we compare the exact value
$E(\alpha)$, the result of the optimized expansion $E_1(\alpha
,\alpha_1^{PMS})$ and the results of the odinary perturbation theory
up to the $n$-th order ($n=0,~1$) {\it i.e.} not expanded values of
$E_0(\alpha )$ and $E_1(\alpha ,\alpha )$.  All the $[i,j]$ Pad\'{e}
approximants $E^{[i,j]}(\alpha)$ with $i+j\le 3$ are in Tab.~3.

We give numerical estimates for three different values of
the parameter $\alpha$ that represent typical cases

\noindent
1. $\alpha =0.1$: the asymptotic expansion is applicable and
justified;

\noindent
2. $\alpha =0.5$: the asymptotic expansion is applicable but
one has to deal with the high orders corrections to achieve
a satisfactory accuracy. An improvement of the perturbation
theory is desirable;

\noindent
3. $\alpha =1.0$: the asymptotic expansion in principle can provides
only $\sim 2\%$ accuracy after summation of $\sim 13$ terms. Then the
terms start to grow.  The perturbation theory must be reformulated.

As we can see a naive attempt to improve the convergence
of the perturbative expansion simply keeping the exact
(not expanded) value of $E_n(\alpha ,\alpha )$  that sums up
some next-to-leading corrections gives a good result but is
essentially insufficient if $\alpha $ is large enough.
This shows that optimization is important for convergence.
Though we cannot directly demonstrate that without the optimization
the series~(\ref{10}) becomes divergent as it takes place in the case
of anharmonic oscillator \cite{5} the numerical analysis
clearly shows an advantage of optimized expansion.
Indeed in all these cases the best convergence is achieved within
OPT.  Even for  $\alpha =1.0$ taking only
the first order correction we reach $\sim 2\%$ accuracy.  Using the
PMS prescription we can also correctly estimate an error of the
result.  For example for $\alpha =1.0$ the naive estimate is
$(E^{(1)}(1.0,\alpha^{PMS})/E^{(0)}(1.0,\alpha^{PMS}))^2\sim 0.02$
that coincides with the real uncertainty (see Tab.~1).

We should note that in all the cases the terms of $\alpha$
expansion taken into account in our numerical analysis are far from
the critical point where the series begins to diverge. The bad
convergence of the series reveals only in the fact that they decrease
quite slow.  On the other hand even the third order correction is
hardly available within ordinary perturbation theory.  So the
accuracy of the perturbation theory is restricted rather by technical
reasons than by asymptotic character of the series.  Thus
the optimization is not only a resummation prescription
that is useful in high orders of asymptotic expansion but also
gives an opportunity to improve accuracy of perturbation theory in
low orders.  This is an important benefit especially for non-trivial
systems where high orders calculations are impossible.

A remark about Pad\'{e} approach is in order.
As we can see some Pad\'{e} approximants are closer to the
exact result  than the plain  asymptotic expansion.
However it is not possible to make a choice between various
approximants until the exact result  or the general structure of the
series are known. Moreover in high orders where the asymptotic
character of the series reveals the Pad\'{e} theory becomes useless.
The matter is that Pad\'{e} approximants because of their specific
structure can correctly reproduce only pole-like singularity
while the function $E(\alpha)$ has a branching
point at $\alpha =0$.

\section{The positronium bound state}

The theoretical predictions for the orthopositronium width
reads \cite{2}

$$
\Gamma_{o-Ps}=
m\alpha ^6{(2\pi^2-18)\over 9\pi}\left[ 1-
10.282\left({\alpha \over\pi}\right)+\right.
$$
\begin{equation}
\left. +{1\over 3}\alpha ^2\ln\alpha +B
\left({\alpha \over\pi}\right)^2+\ldots
\right].
\label{20}
\end{equation}
where the coefficient $B$ has not yet been computed but it
is expected to be about 250 to bring theory and experiment
into agreement. The coefficients of eq.~(\ref{20}) tend to
grow rapidly. In spite of the value $\alpha =1/137$ seems to be
small enough the contribution of the second order term is about
$0.2\%$ of the leading one. This resembles the behavior of the
asymptotic series in our toy model with $0.1<\alpha<0.5$
where the asymptotic expansion can, in principle, provide a sufficient
accuracy but it requires to compute high order terms that
is not possible by pure technical reasons.  So we hope that after
optimization of the expansion an agreement between experimental and
theoretical values will be reached already in the first order of
OPT.

The main problem of  OPT is to find an
appropriate form of "unperturbed" action which, on the one hand
reflects the main properties of the exact theory and, on the other
hand does not lead to extremely cumbersome calculations. The most
direct way for the positronium bound state is as follows.
By now there are two general methods to develop
the systematic perturbation theory for the positronium:
the non-relativistic $1/c$ expansion \cite{st} and the method based on
Bethe-Salpeter equation  \cite{bs}. The
non-relativistic Coulomb solution and Barbieri-Remiddi solution
\cite{br} which are used to built the leading order approximations
for the positronium bound state Green's function  within these
approaches involve the physical fine structure constant as a
parameter.  In analogy with our toy model one can replace it with a
changeable parameter $\alpha '$.  Then in every order of the new
perturbation theory one has to keep the exact dependence on $\alpha
'$ and fix it using some optimization prescription. We should
emphasize that the optimized value has no direct relation to the
physical fine structure constant.  This is an auxiliary parameter and
constructing the OPT, for example, for
orthopositronium bound energy we would obtain a different value of
this parameter. Unlike the case of the asymptotic $\alpha $ expansion
two above start approximations can lead to the different results if
one is interested in the optimized expansion but it is not obvious
which is preferable.
Though the detail analysis of the positronium is a  non-trivial
technical problem that is a subject of a separate publication.

To conclude we should note that an attempt to improve the perturbation
theory in orthopositronium width analysis using Pad\'{e} approximants has
been made in ref.~\cite{pd}.  However it looks artificially because an
information about structure of the asymptotic expansion~(\ref{20}) is
absent. On the other hand  OPT seems to be the
most appropriate tool to deal with such a problem because
it speeds up the convergence of the perturbation
series choosing the most  natural  start approximation for a {\it
specific} model not by an extraneous mathematical trick.
This is an automatic summation device that need not an input information
on the form of the singularity of the bound state Green's function
in the coupling constant but reproduce it via optimization.  This
feature can be merely observed in our toy model where the form of the
singularity of the ground state energy at $\alpha =0$ in the
auxiliary and original theories are essentially different but
OPT reproduces the correct singular
$\alpha$ dependence that is absorbed by the optimized value of
auxiliary parameter $\alpha '$.  This general property of optimized
perturbation theory allows to cope even with Borel non-summable
series \cite{5} while a class of the problems where
the Pad\'{e} theory can be successfully applied is quite restricted.

\vspace{0.5cm}
\noindent
{\large \bf Acknowledgments}

\noindent
This work is supported in part
by Russian Fund for Fundamental Research under
Contract No. 93-02-14428 and  by Soros
Foundation. The work of A.A.Penin has been made possible by a
fellowship of Royal Swedish Academy of Sciences and is carried out
under the research program of International Center for Fundamental
Physics in Moscow.

\vspace{0.5cm}
\noindent
{\large \bf Figure Captions}

\noindent
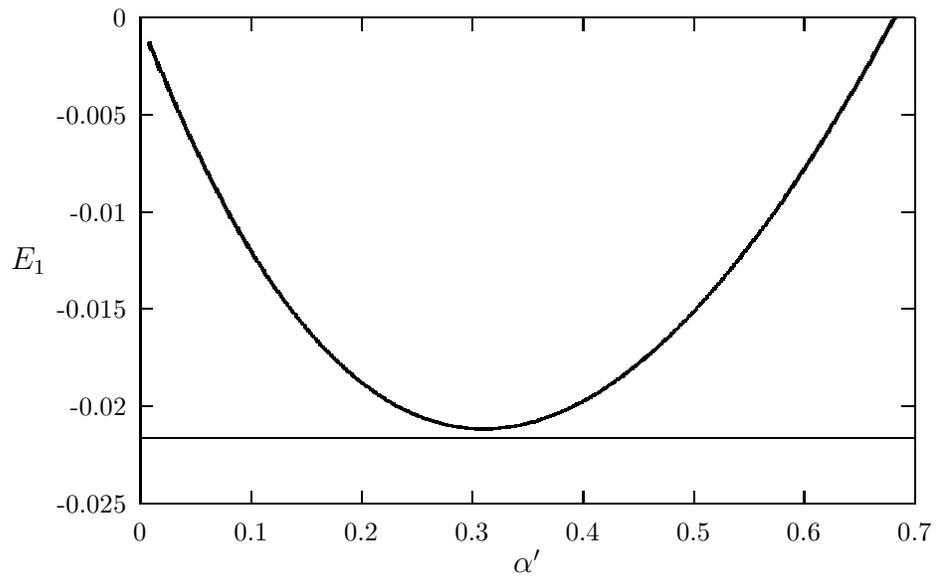
Fig. 1.  The curve line depicts the function  $E_1(1.0,\alpha ')$.
The straight line corresponds to the exact value $E(1.0)$.

\noindent
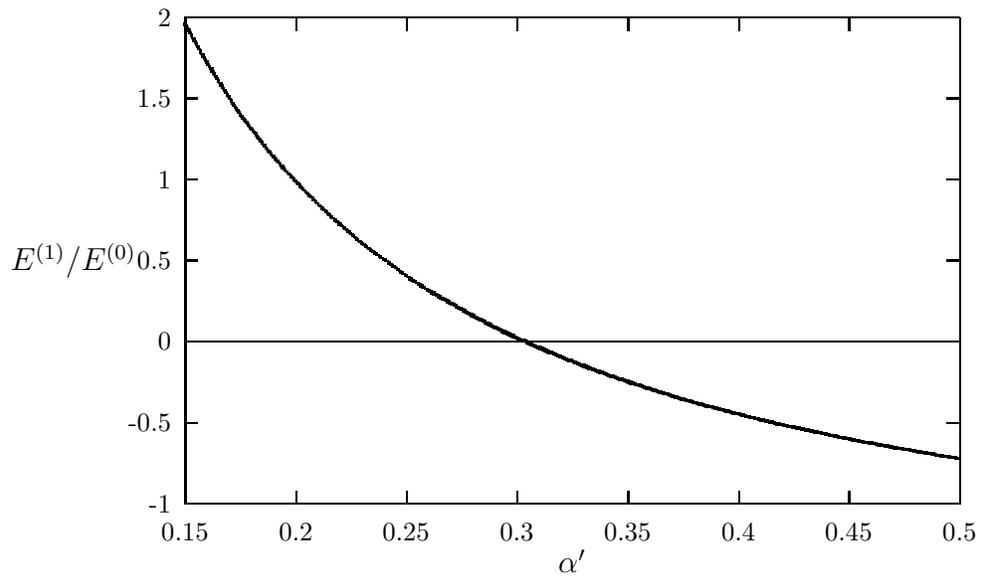
Fig. 2. The function $E^{(1)}(1.0,\alpha ')/E^{(0)}(1.0,\alpha ')$.


\newpage

{\Large \bf Tables}

\vspace{5mm}

\noindent
Table 1. The exact bound energy $E(\alpha)$ and the results of
the asymptotic expansion up to the $n$-th order
$\tilde E_n(\alpha)$ $(n=0,~1,~2,~3)$.
\begin{center}
\begin{tabular}{|c|c|c|c|c|c|}
\multicolumn{6}{c}
{Table~1.}
\\  \hline
$\alpha$ & $E(\alpha)$ & $\tilde E_0(\alpha)$ & $\tilde E_1(\alpha)$ & $\tilde
E_2(\alpha)$ & $\tilde E_3(\alpha)$ \\ \hline
0.1      & 0.2101      & 0.2500                 & 0.2074             & 0.2105
            & 0.2104               \\ \hline
0.5      & 2.426       & 6.250                  & 0.9292             & 2.898
            & 2.284                \\ \hline
1.0      & 2.144       & 25.00                  & -17.57             & 13.94
            & -5.721               \\ \hline
\end{tabular}
\end{center}

\vspace{5mm}

\noindent
Table 2. The exact bound energy $E(\alpha)$, the result of the optimized
expansion up to the  first order $E_1(\alpha,\alpha_1^{PMS})$
and the results of the odinary perturbation theory up to the $n$-th
order $E_n(\alpha,\alpha)$ $(n=0,~1)$.
\begin{center}
\begin{tabular}{|c|c|c|c|c|}
\multicolumn{5}{c}
{Table 2.}
 \\  \hline
$\alpha$ & $E(\alpha)$ & $E_1(\alpha,\alpha_1^{PMS})$ & $E_0(\alpha)$ &
$E_1(\alpha,\alpha)$  \\ \hline
0.1      & 0.2101      & 0.2104                       & 0.2430        & 0.2092
              \\ \hline
0.5      & 2.426       & 2.424                        & 5.531         & 1.915
              \\ \hline
1.0      & 2.144       & 2.116                        & 20.13         & -4.200
              \\ \hline
\end{tabular}
\end{center}

\vspace{5mm}

\noindent
Table 3.The exact bound energy $E(\alpha)$ and various $[i,j]$
Pad\'{e} approximants $E^{[i,j]}(\alpha)$ with $i+j\le 3$.
\begin{center}
%

\begin{center}
{\large \bf Fig. 2}
\end{center}

\end{document}